\documentclass[aps,prd,10pt,nofootinbib,twocolumn,eqsecnum,showpacs,showkeys,superscriptaddress,preprintnumbers,altaffilletter]{revtex4-1}

\usepackage{graphicx}
\usepackage{dcolumn}
\usepackage{amssymb}
\usepackage{amsmath}
\usepackage{amsfonts}
\usepackage{amsbsy}
\usepackage{color}
\usepackage{rotating}
\usepackage[english]{babel}
\usepackage[utf8]{inputenc}

\def\be{\begin{equation}}
\def\ee{\end{equation}}

\def\bea{\begin{eqnarray}}
\def\eea{\end{eqnarray}}

\begin{document}

\title{$f(R)$ gravity modifications: from the action to the data}

\author{Ruth Lazkoz}
\email{ruth.lazkoz@ehu.eus}
\affiliation{Department of Theoretical Physics, University of the Basque Country UPV/EHU, P.O. Box 644, 48080 Bilbao, Spain}
\author{Mar\'ia Ortiz-Ba\~nos}
\email{maria.ortiz@ehu.eus}
\affiliation{Department of Theoretical Physics, University of the Basque Country UPV/EHU, P.O. Box 644, 48080 Bilbao, Spain}
\author{Vincenzo Salzano}
\email{vincenzo.salzano@usz.edu.pl}
\affiliation{Institute of Physics, Faculty of Mathematics and Physics, University of Szczecin, Wielkopolska 15, 70-451 Szczecin, Poland}


\begin{abstract}
It is a very well established matter nowadays that many modified gravity models can offer a sound alternative to General Relativity for the description of the accelerated expansion of the universe. But it is also equally well known that no clear and sharp discrimination between any alternative theory and the classical one has been found so far. In this work, we attempt at formulating a different approach starting from the general class of $f(R)$ theories as test probes: we try to reformulate $f(R)$ Lagrangian terms as explicit functions of the redshift, i.e., as $f(z)$. In this context, the $f(R)$ setting to the consensus cosmological model, the $\Lambda$CDM model, can be written as a polynomial including just a constant and a third-order term. Starting from this result, we propose various different polynomial parameterizations $f(z)$, including new terms which would allow for deviations from $\Lambda$CDM, and we thoroughly compare them with observational data.
 While on the one hand we have
found no statistically preference for our proposals (even if some of them are as good as
$\Lambda$CDM by using Bayesian Evidence comparison), we think that our novel approach could provide a
different perspective for the development of new and observationally reliable alternative models of
gravity.
\end{abstract}

\maketitle

\section{Introduction}

Almost $20$ years have passed since distant Ia supernovae showed for the first time that the universe is expanding in an accelerated way \cite{riess,perl,Perlmutter}. Since then, this evidence has been supported by many other observations as cosmic microwave background anisotropies (CMB) \cite{Planck} and large scale structure \cite{sdss1,Blake:2012pj,Parkinson:2012vd,Kazin:2014qga,sdss2,Beutler:2015tla} but the question of the origin of this acceleration is still waiting for an answer. Despite the favorable experimental results which support the standard paradigm \cite{Planck,Aubourg:2014yra}, the $\Lambda$CDM model $-$ a cosmological constant \cite{Carroll:1991mt,Sahni:1999gb,Peebles:2002gy} plus cold dark matter in the framework of general relativity (GR) $-$ there are some important theoretical shortcomings in it as well as tensions between data and theory \cite{beyond}, the latest one being the discrepancy \cite{Bernal2016} between the values of the Hubble constant $(H_0)$ as measured by \textit{Planck} \cite{Planck} (i.e. assuming $\Lambda$CDM as the background cosmology) and the local measurements from Cepheids \cite{Riess2016}.

All these facts raise the need for the formulation of many other approaches and theories. Many authors do not leave the realm of GR, and explore dynamical alternatives to the cosmological constant, the so-called dark energy models, where a new component of the mass-energy tensor is added. These alternative settings can be accomplished in the most varied ways, from the theoretical side \cite{Frieman:2008sn,Li:2011sd} to the phenomenological one, where the most common approach is a generalization of the cosmological constant itself by making the equation of state have additional parameters and/or geometry dependence \cite{Cooray:1999da,Efstathiou:1999tm,Astier:2000as,Goliath:2001af,Chevallier,Linder,Weller:2001gf,Sahni:2002fz,Padmanabhan:2002vv,Wetterich:2004pv,Jassal:2004ej,Komatsu:2008hk,Wang:2008vja,Ma:2011nc,Barboza:2011py,Sendra:2011pt}. 
Many others try to give up on GR itself and propose entirely new alternative theories of gravity, also known as modified gravity theories. Even in this case, the number of ways in which GR can be extended and/or modified is extremely large; for a non-exhaustive list see, for example, \cite{Berti:2015itd,Review}, but the border between the dark energy and the modified gravity formulation is not clearly paved \cite{Sotiriou:2007zu,Tsujikawa:2010sc,Tsujikawa:2010zza,Joyce:2016vqv}.

Among the plethora of alternative models proposed, in this work we will focus on the so-called $f(R)$ framework, also known as fourth-order theories or extended theories of gravity \cite{Nojiri:2006ri,Capozziello:2007ec,faraoni,Capozziello:2011et,Capozziello:2015lza}. The easiest and most basic approach to this kind of theories consists in replacing the Ricci scalar, $R$, appearing in the Hilbert-Einstein action of GR with a general function $f(R)$. 
The $f(R)$ proposals, of course, cannot be arbitrary, but they have to be able to both fit the cosmological data and to satisfy the Solar System constraints, given that on such scales GR has been experimentally tested and confirmed. Some examples of such viable models are in \cite{star,hu,Appleby:2007vb,Nojiri:2007jr,Nojiri:2007as,Tsujikawa:2007xu}. These theories have also been intensively analyzed, by comparing them with cosmological probes \cite{Capozziello:2003gx,Capozziello:2004vh,Capozziello:2008qc,BouhmadiLopez:2010pp,Cataneo:2014kaa,Basilakos:2013nfa,Dev:2008rx,Nunes:2016drj}, with the internal dynamics of many different-scale gravitational structures \cite{Capozziello:2004us,Capozziello:2006uv,Capozziello:2006ph,Capozziello:2006dp,Capozziello:2008ny,Capozziello:2009ka,Lin:2010hk}, and with cosmological simulations \cite{Oyaizu:2008sr,Oyaizu:2008tb,Schmidt:2008tn,Zhao:2010qy,Li:2012by,Arnold:2013nfr,Arnold:2014qha}, eventually putting eather too weak or too severe (i.e. very consistent with the $\Lambda$CDM limits) constraints on their viability.

The most common approach when one tries to accomplish this task is to first propose an $f(R)$ expression at the level of the Lagrangian, with the requirement that it satisfies some priors (e.g. Solar System constraints), then to solve the corresponding dynamical equations, and finally to test them against the data. The main obstacle in this procedure is that the fourth order differential equations which come out from a very general $f(R)$ Lagrangian are not analytically solvable. For cosmological applications, consider that the Friedmann equations can result to be quite complicated third order differential equation in $H$ \cite{saez2}, the expansion rate function which is generally needed in order to calculate cosmological distances, and we might miss an easy analytical expression for it (see \cite{delaCruz-Dombriz:2015tye} for an application to the famous model from \cite{hu}). In order to overcome such difficulties, thus, eather one needs to fix a functional form for the $f(R)$ function in order to obtain analytically manageable equations \cite{Capozziello:2003gx}, or some phenomenological ans\"{a}tze for an $H$ compatible with the given $f(R)$ have to be proposed \cite{capo}. Alternatively, one can go the other way around, that is, to recover analytically a specific $f(R)$ model from any requested standard cosmology \cite{Nojiri:2009kx} or given any $H(t)$ proposal \cite{delaCruz-Dombriz:2015tye, saez2}, or to reconstruct numerically the $f(R)$ from the data \cite{Capozziello:2005ku}. It has not to be forgotten, on the other hand, that for $f(R)$ theories a scalar-tensor equivalence holds \cite{Capozziello:2005mj}, meaning that a scalar field can be introduced, coupled to gravity (geometrically described by a standard $f(R)=R$) and following a given potential, which is completely equivalent to the given $f(R)$ model.

In all the previous cases, anyway, there are some flaws, or limitations: in one way or another, we need to make assumptions at some step, so that the results will be always somewhat biased by these choices and will not be as general as they should or as we would like; in some cases, the relation between $f(R)$ and $H$ is generally approximate and some information might be lost (especially if advocating the scalar-tensor equivalence). One usual approach is to take into account that the $f(R)$ cosmology should mimic the $\Lambda$CDM model as much as possible given that, at the end of the day, this model gives the best description to most of the cosmological data available nowadays. Thus, the $f(R)$ cosmology should recover a matter dominated stage at high redshift, and should show accelerated expansion at low redshift, ideally, without a true cosmological constant, but through purely geometrical terms. Apart from setting some general limits based on these considerations, there really is not much more information one can provide a priori in order to yield sensible $f(R)$ theories.

Our present study is precisely related to this topic. We want to explore this problem but using a different approach, a sort of \textit{practical poor cosmologist} approach. The vast amount of probes in cosmology provides us with a big variety of measurements usually corresponding to specific values of the redshift; thus, we think it would be interesting to have gravity models which are described in terms of this observable, so that we are provided with a more straightforward way to test the models against the observations. Studies exist in the literature where the redshift formulation appears \cite{capo}, however they are not focused on constructing a sensible $f(z)$ model but rather on proposing ans\"{a}tze both for the Hubble function $H$ and for the $f(R)$ function, and then fitting the parameters in order to do a cosmographic reconstruction. Our present analysis will seek a way to provide reasonable $f(z)$ models since the beginning, i.e., from the action, so that at a later stage one can numerically solve the dynamics of the Universe, to test their validity and study their deviation with respect to the $\Lambda$CDM scenario.

We begin with Sec.~\ref{sec:introducing} by introducing briefly a very general approach to $f(R)$ gravity, with a description of the modified Friedmann equations and how we build $f(R)$ as a function of the redshift $z$. The expression of the derivatives of $f(R)$ with respect to $R$, and of $R$ with respect to time are provided in terms of derivatives with respect to the redshift $z$. We then consider various different dark energy and modified gravity scenarios and calculate the corresponding high and low redshift limits to validate the choice of our proposals. In Sec.~\ref{sec:data} we describe the observational data used in our analysis; in Sec.~\ref{sec:results} we discuss the obtained results and in Sec.~\ref{sec:conclusions} we give some conclusions.

\section{From $\boldsymbol{f(R)}$ to $\boldsymbol{f(z)}$}\label{sec:introducing}

We consider the most general $f(R)$-type modification to the Einstein-Hilbert action described by the action \cite{faraoni}
\be
S=\int d^4x\sqrt{-g}[f(R)+\mathcal{L}_m],
\ee
where $g$ is the determinant of the metric $g_{\mu\nu}$ and $\mathcal{L}_m$ is the Lagrangian of any considered energy-matter component. In order to obtain the field equations one has to vary the action with respect to the metric field, $g_{\mu\nu}$, ending up with
\be
R_{\mu\nu}f_R-\frac{1}{2}g_{\mu\nu}f+(g_{\mu\nu}\Box-\nabla_{\mu}\nabla_{\nu})f_R=T_{\mu\nu}^{m},
\ee
where $R_{\mu\nu}$ is the Ricci tensor, $\Box$ and $\nabla$ are respectively the d'Alembertian and Laplacian operators, $T_{\mu\nu}^{m}$ is the stress energy tensor, and we define $f_R\equiv df/dR$. In the remainder of the paper we assume a background Friedmann-Lema\^itre-Robertson-Walker (FLRW) metric in spherical coordinates
\be
ds^2=-c^2dt^2+a(t)^2\left[\frac{dr^2}{1-k r^{2}}+r^2(d\theta^2+\sin^2\theta d\phi^2)\right],
\ee
with $c$, the speed of light in vacuum and $a(t)$, the scale factor, and we restrict our considerations to spatially flat spaces ($k=0$), with matter and radiation as the only contribution to the stress-energy tensor. Thus we will consider $f(R)$ as a purely geometrical contribution even if, as stated before, we can always find a scalar field, entering the stress-energy tensor, and completely equivalent to the original proposed $f(R)$. Finally, we obtain the generalized Friedmann equations which govern the dynamics of the universe at large scales, namely \cite{saez2}
\bea
H^2=\frac{1}{3f_R}\left(\rho_m+\rho_r+\frac{Rf_R-f}{2}-3H\dot{R}f_{2R}\right)\label{fried},\\
-3H^2-2\dot{H}=\frac{1}{f_R}\left[\dot{R}^2 f_{3R}+\left(2H\dot{R}+\ddot{R}\right)f_{2R}\right.\nonumber\\
\left.+\frac{1}{2}(f-Rf_R)\right],
\eea
where $\cdot\equiv d/dt$ and we have defined $f_{2R} \equiv df^{2}/dR^{2}$, $f_{3R} \equiv df^{3}/dR^{3}$, together with the energy conservation equation for standard energy-matter perfect fluids:
\begin{equation}
\dot{\rho_{X}}(t) + 3 H(t) \left[ \rho_{X}(t) + 3 \frac{p_{X}(t)}{c^2}\right] = 0\;,
\end{equation}
where the suffix $X$ stands for matter, radiation or any fluid in the stress-energy tensor.

For our goals, the next step is to perform a change of variables in order to have all the derivatives appearing in Eq.~(\ref{fried}) in terms of the redshift. First, we note that by combining the Friedmann equations we obtain
\begin{equation}
R =-3 (H^2)_z (1+z)+12H^2\label{ric}\; ,
\end{equation}
which is the usual definition for the Ricci scalar for homogeneous and isotropic flat FLRW spacetimes. From this, we calculate
\begin{eqnarray}
R_{z}&=& 9 (H^2)_z -3 (1+z) (H^2)_{2z}\label{ric2}, \\
R_{2z}&=& 6 (H^2)_{2z} -3 (1+z) (H^2)_{3z}. \nonumber
\end{eqnarray}
where the subindex $z$ means derivative w.r.t. the redshift. Finally, one gets:
\begin{eqnarray}
f_R&=&R_z^{-1}f_z, \nonumber \\
f_{2R}&=&(f_{2z}R_{z}-f_zR_{2z})R_z^{-3}, \nonumber \\
f_{3R}&=&\frac{f_{3z}}{R^3_z}-\frac{f_zR_{3z}+3f_{2z}R_{2z}}{R^4_z}+\frac{3f_zR_{2z}^2}{R_z^5}, \\
\dot{R}&=&-(1+z)HR_z ,\nonumber \\
\ddot{R}&=&(1+z)H[HR_z+(1+z)(H_zR_z+HR_{2z})] .\nonumber
\end{eqnarray}
In this work we will provide $f(R)$ as $f(z)$, we will solve Eq.~(\ref{fried}) numerically for $H$, and we will compare it to observational data. In terms of the redshift, the first Friedmann equation Eq.~(\ref{fried}) now reads:
\begin{equation}
H^2=\frac{\rho_{f}}{3}+\frac{R_z \left[(1+z)^4\Omega_r+\Omega_m (1+z)^3\right]}{f_z}\label{friedmann},
\end{equation}
with
\begin{equation}
\rho_{f}=\frac{R_z}{f_z} \left[ \frac{1}{2}\left(\frac{R f_z}{R_z}-f\right)+\frac{3(1+z) H^2 (R_z f_{2z}-R_{2z} f_z)}{R_z^2}\right].
\end{equation}
Clearly, here we have a third order differential equation, for which we need to set initial conditions for $H$, $H_z$ and $H_{2z}$; we will discuss in the next section how we choose such initial conditions. 

\subsection{Requirements for $f(z)$ proposals}

Once we have the equation for $H$, the next step is to provide a ``good'' $f(z)$ model to test against data. We know that a spatially flat $\Lambda$CDM universe can be expressed as the $f(R)$ theory:
\be\label{gr}
 f_{\Lambda}(R)=R-2\Lambda.
\ee
If we assume the universe filled wth matter and radiation, then the first Friedmann equation can be cast into the following form:
\be
 H^2(z)=\Omega_m(1+z)^3+\Omega_r(1+z)^4+(1-\Omega_m-\Omega_r)\, ,
\ee
where we have defined the dimensionless density parameters as
\begin{equation}
\Omega_{i} = \frac{8\pi G \rho_{i}}{3 H^{2}_{0}}\; ,
\end{equation}
with $i=m,r$ indicating, respectively, matter and radiation, and $1-\Omega_m-\Omega_r$ corresponding to the cosmological constant. To get $f_{\Lambda}$ in terms of the redshift we can use  Eqs.~(\ref{ric}), thus obtaining
\begin{eqnarray}
R(z)&=& 12  (1 - \Omega_m-\Omega_r)  + 3  \Omega_m (1 + z)^3\label{rl2},\\
f_{\Lambda}(z)&=& 6  (1 - \Omega_m-\Omega_r)  + 3  \Omega_m (1 + z)^3.\label{fl}
\end{eqnarray}
Here we have normalized by $H_0^2$, the Hubble constant $H_{0}\equiv H(z=0)$, just to simplify notation, and we will keep this notation for the whole analysis.

Taking this into account we would like to choose an $f(z)$ which is somehow the simplest generalization of the latter expression, that is, a polynomial with more terms and not only a constant and a third-order one. In order to set some restrictions when choosing a specific model, a useful analysis is to study the high and low redshift limit of $R$ and $f(R)$ for the case of $\Lambda$CDM and other different models of dark energy or modified gravity theories. At the end of the day, even if $\Lambda$CDM may not be the model which really underlies our universe, it is the one which, so far, best describes it. Thus, any generalization, should have a behaviour which should follow its same trends. If we were able to detect some special feature, we could propose a reasonable $f(z)$ according to this.

Actually, Eqs.~(\ref{rl2})~-~(\ref{fl}) exactly provide us with such information; it is straightforward and somehow trivial to verify that the high and low redshift limits of $\Lambda$CDM for both the Ricci scalar and the $f_{\Lambda}(R)$ function (which in this case are identical, because of GR) read
\begin{eqnarray}
\lim_{z\rightarrow\infty}R(z)&=&\lim_{z\rightarrow\infty}f_{\Lambda}(z) = 3\Omega_m(1+z)^3, \\
\lim_{z\rightarrow 0}R(z)&=&\lim_{z\rightarrow 0}f_{\Lambda}(z) =\, const.
\end{eqnarray}

Let us generalize a little bit such scenario, considering the most popular and common dark energy model used in the usual references, the so-called Chevallier-Polarski-Linder \cite{Chevallier,Linder} (CPL) parametrization. In this model the dark energy fluid has a dynamical equation of state which is linear in the scale factor:
\be
w(a)=w_0+w_a(1-a)\, .
\ee
The well-known expression for the first Friedmann equation in this case is:
\begin{eqnarray}
H^2&=&\Omega_m a^{-3}+\Omega_r a^{-4} \\
&+&(1-\Omega_m-\Omega_r) e^{3w_a (a-1)}a^{-3(1+w_0+w_a)}\, , \nonumber
\end{eqnarray}
from this, we can compute the Ricci scalar:
\begin{eqnarray}
R(a)&=&3\Omega_m a^{-3}+3(1-\Omega_m-\Omega_r) \\
&\times&\left(1-3w(a)\right)e^{3w_a (a-1)}a^{-3(1+w_0+w_a)}\, . \nonumber
\end{eqnarray}
As we are still working in the context of general relativity, $f_{CPL} = R$ and the CPL dark energy fluid is included in the stress-energy tensor.
Let us notice that $w(a=1)=w_0$ has to be negative today, in order to lead to the observed accelerated expansion of the universe; and we can further assume the strong prior $w_0+w_a<0$, as confirmed by observations \cite{Planck}, which implies that the universe was matter dominated at early times. Accordingly,
\be
\lim_{z\rightarrow \infty}R(z)=3\Omega_m (1+z)^{3}\, ,
\ee
while for the low redshift limit one finds:
\begin{eqnarray}
\lim_{z\rightarrow 0}R(z)&=&3(1-\Omega_m-\Omega_r)(1-3w_0)+3\Omega_m \\
&=&\; const. \nonumber
\end{eqnarray}
Setting $w_0=-1$ we would get the limit for the $\Lambda$CDM model.

Another possibility is given by the ``early dark energy'' models \cite{ede1,ede2}. We will focus on the model discussed in \cite{ee}.
It is a phenomenological scenario which behaves as a cosmological constant at early times and decays rapidly at late times:
\begin{eqnarray}
H(a)^2&=&\Omega_m a^{-3}+\Omega_r a^{-4}+(1-\Omega_m-\Omega_r-\Omega_{ee})  \nonumber \\
&+&\Omega_{ee}\frac{(1+a_c)^2}{a^6+a_c^6}\; ,
\end{eqnarray}
where $\Omega_{ee}$ is the fractional energy density of the early dark energy today, and $a_c=1/(1+z_c)$ is the critical value of the scale at which it shifts from the early-time behaviour to the late-time behaviour. Computing the Ricci scalar in terms of the redshift, one gets
\begin{eqnarray}
R(z)&=&12(1-\Omega_{ee}-\Omega_m-\Omega_r)+3\Omega_m(1+z)^3~ ~\\
&+&\frac{6\Omega_{ee}(1+a_c)^2}{a_c^6+\frac{1}{(1+z)^6}}\left(2-\frac{3}{1+(1+z)^6a_c^6}\right)\,. \nonumber
\end{eqnarray}
Taking the limits one finds:
\begin{eqnarray}
\lim_{z\rightarrow \infty}R(z)&=&3\Omega_m (1+z)^{3}\\
\lim_{z\rightarrow 0}R(z)&=&12(1-\Omega_{ee}-\Omega_r)-9\Omega_m \\
&+&6\Omega_{ee}\frac{(1+a_c)^2}{1+a_c^6}\left(2-\frac{3}{1+a_c^6}\right)\nonumber \\
&=& \, const. \nonumber
\end{eqnarray}
Thus, in these two cases, we qualitatively recover the same limits as $\Lambda$CDM. While this is somehow expected, because both the CPL parametrization and the early dark energy models are generalizations of the cosmological constant and contain it as a special case, we are also interested in exploring the limits of more radically different approaches. For example, we will consider the Ricci dark energy model \cite{ricci} which belongs to the so-called Holographic Dark Energy models. The basic idea behind this class of models \cite{hde} is that our universe is in a sense finite and can be described by a two dimensional spherical holographic screen, thus there must be finite size effects. At least on a theoretical background, there is a big conceptual difference with respect to a ``simple'' cosmological constant; moreover, another interesting feature here is that $\Lambda$CDM is not explicitly included in this model. This difference, anyway, as it happens in many cases, does not necessarily translate into a difference in the quantitative description of the observational data. In this model the cosmological evolution is governed by
\begin{eqnarray}
H^2&=&\frac{2\Omega_m}{2-\gamma}(1+z)^3+\Omega_r(1+z)^4\nonumber\\
&+& \left(1-\Omega_r-\frac{2\Omega_m}{2-\gamma}\right)(1+z)^{4-\frac{2}{\gamma}}\; ,
\end{eqnarray}
from which the Ricci scalar reads
\begin{eqnarray}
R(z)&=& \frac{6 \Omega_m (1+z)^3}{2-\gamma }   \nonumber \\
& & +6(1+z)^4 A_{\gamma}(z) \left(1-\Omega_r -\frac{2\Omega_m}{2-\gamma} \right),
\end{eqnarray}
with
\be
A_{\gamma}(z) \equiv\frac{(1+z)^{-2/\gamma }}{\gamma} \; .
\ee
One can notice that the first term in $R$ will dominate as far as $\gamma<2$. And, actually, this parameter has been constrained with observational data \cite{chinos} to be $\gamma=0.325 ^{+0.009}_{-0.010}$, so that one can write
\be
\lim_{z\rightarrow \infty}R(z)=\frac{6\Omega_m (1+z)^{3}}{2-\gamma}.
\ee
The low redshift limit exhibits, like in the previous cases, a constant behaviour:
\be
\lim_{z\rightarrow 0}R(z)=\frac{6}{\gamma}\left(1-\frac{2\Omega_m}{2-\gamma}-\Omega_r\right)+\frac{6\Omega_m}{2-\gamma}=const.
\ee

We also consider another well known example of modified gravity, the Dvali-Gabadadze-Porrati (DGP) model \cite{dgp}, in which gravity leaks
off the four dimensional Minkowski brane into the five dimensional bulk Minkowski space-time and such setting should yield a self-acceleration of the universe without introducing dark energy. The dynamics for this model is given by the modified Friedmann equation \cite{Deffayet:2001pu}
\be\label{hdgp}
H(z)=\sqrt{\Omega_m(1+z)^3+\Omega_r(1+z)^4+\Omega_{r_c}}+\sqrt{\Omega_{r_c}}.
\ee
As in the previous case, the $\Lambda$CDM model is not a sub-case within this theory. Plugging Eq.~(\ref{hdgp}) into Eq.~(\ref{ric}) we get:
\begin{eqnarray}
R(z)&=& 3 \Omega_m (1+z)^3+12\sqrt{\Omega_{r_c}}h(z)+24\Omega_{r_c} \nonumber\\
&+&\frac{3\sqrt{\Omega_{r_c}}}{h(z)}\left(4\Omega_{r_c}+\Omega_m (1+z)^3\right),
\end{eqnarray}
with
\be
h(z)\equiv\sqrt{\Omega_{r_c}+\Omega_m (1+z)^3+\Omega_r (1+z)^4}
\ee
and
\begin{equation}
\Omega_{r_c} = \frac{1}{4 r^{2}_{c} H^{2}_{0}}\, ,
\end{equation}
where $r_{c}$ is the cross-over scale that governs the transition between four-dimensional behavior and five-dimensional behavior. One can easily see that as $z\rightarrow\infty$, the biggest contribution corresponds to
\be
\lim_{z\rightarrow \infty}R(z)=3\Omega_m (1+z)^{3},
\ee
while for the low redshift limit we have:
\begin{eqnarray}
\lim_{z\rightarrow 0}R(z)&=&3\Omega_m+12\sqrt{\Omega_{r_c}}h_{\Omega}+24\Omega_{r_c}\\
&+&\frac{3\sqrt{\Omega_{r_c}}}{h_{\Omega}}\left(4\Omega_{r_c}+\Omega_m\right)=const
\end{eqnarray}
with
\be
h_{\Omega}\equiv\sqrt{\Omega_{rc}+\Omega_m+\Omega_r}.
\ee
It is important to note that the four-dimensional part of the total DGP action has $f_{DGP}(R) = R$. A generalization of this model can be found in \cite{BouhmadiLopez:2009db}.

\subsection{Final proposals}

Given the previous examples and reminding that for all of them we have $f(R)=R$, we can see that in many relevant cases in the literature these hold:
\bea\label{hz}
\lim_{z\rightarrow\infty}R(z)&\propto&(1 + z)^3\, ,\\
\lim_{z\rightarrow \infty}f(z)&\propto& (1+z)^3\, ,\\
\lim_{z\rightarrow0}R(z)&\propto&\, const.\, ,\\
\lim_{z\rightarrow 0}f(z)&\propto&\, const\, ;
\eea
then we can conclude that it is a reasonable choice to propose polynomial expressions with a maximum third-order term as extensions of the $f_{\Lambda}(R)$. Anyway, one could ask what would happen with more general $f(R)$ models alternative to those described above. For example, in \cite{Capozziello:2004vh}, the authors consider the model
\begin{equation}
f(R) = \beta\, R^{n} \;.
\end{equation}
Following \cite{Capozziello:2004vh} it is easy to deduce that
\begin{equation}
R \propto (1+z)^{3/n} \; ,
\end{equation}
from which we have
\begin{equation}
f(R) \propto (1+z)^3 \; .
\end{equation}
This result is not in contrast with previous high redshift limits. Again, in \cite{Capozziello:2004vh}, another $f(R)$ model is considered which given by
\begin{equation}
f(R) = \alpha \, \ln R\, .
\end{equation}
Again, we obtain
\begin{eqnarray}
R &\propto& -\frac{1}{2 \left(9 A (1+z)^3+4\right)^2} \left\{ 3 (9 A+4) \exp^{\frac{3}{2} A \left[(1+z)^3-1\right]} \right. \nonumber \\
&\times& \left. \left[9 A (1+z)^3 \left(9 A (1+z)^3-10\right)-32\right]\right\} \, ,
\end{eqnarray}
with $A = \Omega_{m} H^{2}_{0} \alpha^{-1}$. In this case,
\begin{equation}
\lim_{z\rightarrow\infty}R(z) \propto \exp^{(1+z)^3-1} \; ,
\end{equation}
which implies that
\begin{equation}
\lim_{z\rightarrow\infty}f(R) \propto (1+z)^3 \, .
\end{equation}
Thus, again, we have the same high redshift limit. Note also that in both the $f(R)$ models we have just considered, the low redshift limit is not a proper constant, which means, they do not include a cosmological constant. Of course, we cannot check here all possible models, as this would be out of the purpose of this work, and would be a gigantic, yet useless, exercise, also because even the more commonly used, as the one in \cite{hu}, can not have an analytical solution for $H$ \cite{delaCruz-Dombriz:2015tye}.

But it is clear, that the limits we have described so far, clearly depict a possible trend which we use as guideline for our proposals. Thus, we are going to choose different models to see how much each model deviates with respect to $\Lambda$CDM. Let us notice that some of the models will contain a constant term (so they might resemble a $\Lambda$CDM model) and others will not. The latter may have more interest if we want to provide fully alternative theories to the standard model of cosmology. Finally and all in all, the models we have decided to focus on are:
\begin{enumerate}
 \item $f_0+f_3(1+z)^3$
 \item $f_0+f_1(1+z)+f_2(1+z)^2+f_3(1+z)^3$
 \item $f_0+f_2(1+z)^2+f_3(1+z)^3$
 \item $f_0+f_1(1+z)+f_3(1+z)^3$
 \item $f_{12}(1+z)^{1/2}+f_3(1+z)^3$
 \item $f_{12}(1+z)^{1/2}+f_1(1+z)+f_2(1+z)^2+f_3(1+z)^3$
 \item $f_{14}(1+z)^{1/4}+f_3(1+z)^3$
 \item $f_{14}(1+z)^{1/4}+f_1(1+z)+f_2(1+z)^2+f_3(1+z)^3$
\end{enumerate}

\section{Data}\label{sec:data}

We use the combination of various current observational data to constrain the $f(R)=f(z)$ models described previously. In this section, we describe the
cosmological observations used in this work. We will only consider the observational data related to the expansion history of the universe, i.e., those describing the distance-redshift relations. Specifically, we use the Type Ia Supernovae (SNeIa), the cosmic microwave background (CMB) distance priors, the Baryon Acoustic Oscillations (BAO) data, the expansion rate data from early-type galaxies (ETG), plus a prior on $H_0$.

\subsection{Hubble data}

We use a compilation of Hubble parameter measurements estimated by the differential evolution of passively evolving early-type galaxies used as cosmic chronometers, in the redshift range $0<z<1.97$, and recently updated in \cite{Moresco15}. The corresponding $\chi^2_{H}$ estimator is defined as
\begin{equation}\label{eq:hubble_data}
\chi^2_{H}= \sum_{i=1}^{24} \frac{\left( H(z_{i},\boldsymbol{\theta})-H_{obs}(z_{i}) \right)^{2}}{\sigma^2_{H}(z_{i})} \; ,
\end{equation}
with $\sigma_{H}(z_{i})$ the observational errors on the measured values $H_{obs}(z_{i})$, $\boldsymbol{\theta}$ the vector of the cosmological background parameters. Moreover, we will add a gaussian prior, derived from the Hubble constant value given in \cite{hh}, $H_{0} = 69.6 \pm 0.7$.

\subsection{Type Ia Supernovae data}

We used the SNeIa data from the JLA (Joint-Light-curve Analysis) compilation \cite{Betoule}. This set is made of $740$ SNeIa obtained by the SDSS-II (Sloan Digital Sky Survey) and SNLS (Supenovae Legacy Survey) collaborations, covering the redshift range $0.01<z<1.39$. The $\chi^2$ in this case is defined as
\begin{equation}
\chi^2_{SN} = \Delta \boldsymbol{\mathcal{F}}^{SN} \; \cdot \; \mathbf{C}^{-1}_{SN} \; \cdot \; \Delta  \boldsymbol{\mathcal{F}}^{SN} \; ,
\end{equation}
with $\Delta\boldsymbol{\mathcal{F}} = \mathcal{F}_{theo} - \mathcal{F}_{obs}$, the difference between the observed and the theoretical value of the observable quantity for SNeIa, the distance modulus; and $\mathbf{C}_{SN}$ the total covariance matrix (for a discussion about all the terms involved in its derivation, see \cite{Betoule}). The predicted distance modulus of the SNeIa, $\mu$, given the cosmological model and two other quantities, the stretch $X_{1}$ (a measure of the shape of the SNeIa light-curve) and the color $\mathcal{C}$, is defined as
\begin{equation}\label{eq:m_jla}
\mu(z,\boldsymbol{\theta}) = 5 \log_{10} [ D_{L}(z, \boldsymbol{\theta}) ] - \alpha X_{1} + \beta \mathcal{C} + \mathcal{M}_{B} \; ,
\end{equation}
where $D_{L}$ is the luminosity distance given by
\be
D_L(z,\theta_c)=\frac{c}{H_0}(1+z)\int_{0}^{z}\frac{dz'}{E(z')}
\ee
with $E(z)=H(z)/H_0$ (following \cite{Betoule}, only for SNeIa analysis we assume $H_{0} = 70$ km/s Mpc$^{-1}$) and $c$ the speed of light measured here and now. In this case the vector $\boldsymbol{\theta_{b}}$ will include cosmologically-related parameters and three other fitting parameters: $\alpha$ and $\beta$, which characterize the stretch-luminosity and color-luminosity relationships; and the nuisance parameter $\mathcal{M}_{B}$, expressed as a step function of two more parameters, $\mathcal{M}^{1}_{B}$ and $\Delta_{m}$:
\begin{equation}
\mathcal{M}_{B} = \begin{cases} \mathcal{M}^{1}_{B} & \mbox{if} \quad M_{stellar} < 10^{10} M_{\odot}, \\
\mathcal{M}^{1}_{B} + \Delta_{m} & \mbox{otherwise}\; ,
\end{cases}
\end{equation}
where $M_{stellar}$ is the mass of the host galaxy. Further details are given in \cite{Betoule}.

\subsection{Baryonic Acoustic Oscillations}

The $\chi^2_{BAO}$ for Baryon Acoustic Oscillations (BAO) is defined as
\begin{equation}
\chi^2_{BAO} = \Delta \boldsymbol{\mathcal{F}}^{BAO} \; \cdot \; \mathbf{C}^{-1}_{BAO} \; \cdot \; \Delta  \boldsymbol{\mathcal{F}}^{BAO} \; ,
\end{equation}
where the quantity $\mathcal{F}^{BAO}$ can be different depending on the considered survey. We used data from the WiggleZ Dark Energy Survey, evaluated at redshifts $0.44$, $0.6$ and $0.73$, and given in Table~1 of \cite{Blake:2012pj}; in this case the quantities to be considered are the acoustic parameter
\begin{equation}\label{eq:AWiggle}
A(z, \boldsymbol{\theta}) = 100  \sqrt{\Omega_{m} \, h^2} \frac{D_{V}(z,\boldsymbol{\theta})}{c \, z} \, ,
\end{equation}
and the Alcock-Paczynski distortion parameter
\begin{equation}\label{eq:FWiggle}
F(z, \boldsymbol{\theta}) = (1+z)  \frac{D_{A}(z,\boldsymbol{\theta})\, H(z,\boldsymbol{\theta_{b}})}{c} \, ,
\end{equation}
where $D_{A}$ is the angular diameter distance
\begin{equation}\label{eq:dA}
D_{A}(z, \boldsymbol{\theta_{b}} )  = \frac{c}{H_{0}} \frac{1}{1+z} \ \int_{0}^{z} \frac{\mathrm{d}z'}{E(z',\boldsymbol{\theta})} \; ,
\end{equation}
and $D_{V}$ is the geometric mean of the physical angular diameter distance $D_A$ and of the Hubble function $H(z)$, and defined as
\begin{equation}\label{eq:dV}
D_{V}(z, \boldsymbol{\theta} )  = \left[ (1+z)^2 D^{2}_{A}(z,\boldsymbol{\theta}) \frac{c \, z}{H(z,\boldsymbol{\theta})}\right]^{1/3}.
\end{equation}
We have also considered the data from the SDSS-III Baryon Oscillation Spectroscopic Survey (BOSS) DR$12$, described in \cite{Alam:2016hwk} and expressed as
\begin{equation}
D_{M}(z) \frac{r^{fid}_{s}(z_{d})}{r_{s}(z_{d})} \qquad \mathrm{and} \qquad H(z) \frac{r_{s}(z_{d})}{r^{fid}_{s}(z_{d})} \, ,
\end{equation}
where $r_{s}(z_{d})$ is the sound horizon evaluated at the dragging redshift $z_{d}$; and $r^{fid}_{s}(z_{d})$ is the same sound horizon but calculated for a given fiducial cosmological model used, being equal to $147.78$ Mpc \cite{Alam:2016hwk}. The redshift of the drag epoch is well approximated by \cite{Eisenstein}
\begin{equation}\label{eq:zdrag}
z_{d} = \frac{1291 (\Omega_{m} \, h^2)^{0.251}}{1+0.659(\Omega_{m} \, h^2)^{0.828}} \left[ 1+ b_{1} (\Omega_{b} \, h^2)^{b2}\right]\; ,
\end{equation}
where
\begin{eqnarray}
b_{1} &=& 0.313 (\Omega_{m} \, h^2)^{-0.419} \left[ 1+0.607 (\Omega_{m} \, h^2)^{0.6748}\right], \nonumber \\
b_{2} &=& 0.238 (\Omega_{m} \, h^2)^{0.223}.
\end{eqnarray}
The sound horizon  is defined as:
\begin{equation}\label{eq:soundhor}
r_{s}(z, \boldsymbol{\theta}) = \int^{\infty}_{z} \frac{c_{s}(z')}{H(z',\boldsymbol{\theta})} \mathrm{d}z'\, ,
\end{equation}
with the sound speed
\begin{equation}\label{eq:soundspeed}
c_{s}(z) = \frac{c}{\sqrt{3(1+\overline{R}_{b}\, (1+z)^{-1})}} \; ,
\end{equation}
and
\begin{equation}
\overline{R}_{b} = 31500 \Omega_{b} \, h^{2} \left( T_{CMB}/ 2.7 \right)^{-4}\; ,
\end{equation}
with $T_{CMB} = 2.726$ K. Finally, we have also added data points from Quasar-Lyman $\alpha$ Forest from SDSS-III BOSS DR$11$ \cite{bao3}:
\begin{eqnarray}
\frac{D_{A}(z=2.36)}{r_{s}(z_{d})} &=& 10.8 \pm 0.4\; , \\
\frac{c}{H(z=2.36) r_{s}(z_{d})}  &=& 9.0 \pm 0.3\; .
\end{eqnarray}

\subsection{Cosmic Microwave Background data}

The $\chi^2_{CMB}$ for Cosmic Microwave Background (CMB) is defined as
\begin{equation}
\chi^2_{CMB} = \Delta \boldsymbol{\mathcal{F}}^{CMB} \; \cdot \; \mathbf{C}^{-1}_{CMB} \; \cdot \; \Delta  \boldsymbol{\mathcal{F}}^{CMB} \; ,
\end{equation}
where $\mathcal{F}^{CMB}$ is a vector of quantities taken from \cite{cmb2}, where \textit{Planck} $2015$ data release is analyzed in order to give the so-called shift parameters defined in \cite{Wang2007}. They are related to the positions of the CMB acoustic peaks which depends on the geometry of the model considered and, as such, can be used to discriminate between dark energy models of the different nature. They are defined as:
\begin{eqnarray}
R(\boldsymbol{\theta}) &\equiv& \sqrt{\Omega_m H^2_{0}} \frac{r(z_{\ast},\boldsymbol{\theta})}{c}, \nonumber \\
l_{a}(\boldsymbol{\theta}) &\equiv& \pi \frac{r(z_{\ast},\boldsymbol{\theta})}{r_{s}(z_{\ast},\boldsymbol{\theta})}\, ,
\end{eqnarray}
where we introduce the baryonic density parameter, $\Omega_b$. As before, $r_{s}$ is the comoving sound horizon, evaluated at the photon-decoupling redshift $z_{\ast}$, given by the fitting formula \cite{Hu1996}:
\begin{equation}{\label{eq:zdecoupl}}
z_{\ast} = 1048 \left[ 1 + 0.00124 (\Omega_{b} h^{2})^{-0.738}\right] \left(1+g_{1} (\Omega_{m} h^{2})^{g_{2}} \right) \, ,
\end{equation}
with
\begin{eqnarray}
g_{1} &=& \frac{0.0783 (\Omega_{b} h^{2})^{-0.238}}{1+39.5(\Omega_{b} h^{2})^{-0.763}}\; , \\
g_{2} &=& \frac{0.560}{1+21.1(\Omega_{b} h^{2})^{1.81}} \, ;
\end{eqnarray}
while $r$ is the comoving distance defined as:
\begin{equation}
r(z, \boldsymbol{\theta} )  = \frac{c}{H_{0}} \int_{0}^{z} \frac{\mathrm{d}z'}{E(z',\boldsymbol{\theta})} \mathrm{d}z'\; .
\end{equation}

\subsection{Monte Carlo Markov Chain (MCMC)}

In order to test the predictions of our theory with the available data, we implement an MCMC code in order to minimize the total $\chi^2$ defined as
\begin{equation}
\chi^{2}= \chi_{H}^{2} + \chi_{SN}^{2} + \chi_{BAO}^{2} + \chi_{CMB}^{2} \, .
\end{equation}
As described in Sec.~\ref{sec:introducing}, we will need to solve a third-order differential equation in $H$, in order to recover this quantity and calculate all the required observational signatures. This means we need to specify the initial condition for $H$, $H_z$ and $H_{2z}$. Contrarily to what is done in some literature, where similar equations are solved assuming that the model behaves as $\Lambda$CDM from $z\rightarrow \infty$ up to some finite high redshift value in order to assure a deviation from $\Lambda$CDM only on the narrow redshift range covered by the data, we have decided to leave more freedom to our models to adjust to observations. In particular, we will fix the ``initial conditions'' for our $H(z)$ only at one single redshift point, with no restrictions to its behaviour for both larger and smaller redshift values. The parameters we need to specify are:
\begin{itemize}
 \item[-] $z_{min}$: this redshift corresponds to the early universe regime, i.e. very high $z$. While in theory one should fix $z=\infty$, due to numerical issues we will take it as finite, and we set $z=10^{10}$;
 \item[-] $z_{max}$: this value corresponds to the present time and it is set to $z=0$. Both $z_{min}$ and $z_{max}$ define the redshift interval where the differential equation is solved;
 \item[-] $z_{pivot}$: this point is used to set some ``initial conditions'' of our differential equations system. In our case we impose a $\Lambda$CDM model on this single point. Note that in order to study possible influences of the initial conditions on the final results, we have performed our calculations for the values $z_{piv}=10,~100$.
\end{itemize}
Moreover, analysing carefully the range of values of the parameters in many simulations (by setting different initial conditions) we have decided to apply the following priors on the free parameters in our approach, i.e.:
\begin{itemize}
 \item[-] $0.25<\Omega_m<0.4$;
 \item[-] $0<\Omega_b<0.1$;
 \item[-] $0.65<h<0.75$;
 \item[-] $f_1<0.1$;
 \item[-] $f_2<0.001$;
 \item[-] $\mathcal{M}_{B}<0$.
\end{itemize}
We have verified a posteriori that these choices are licit and do not induce any further strong restriction onto the parameters.

Finally, in order to set up the reliability of one model against the other, we use the Bayesian Evidence, which is generally recognized as the most reliable statistical comparison tool even if it is not completely immune to problems, like its dependence on the choice of priors \cite{Nesseris:2012cq}. We calculate it using the algorithm described in \cite{Mukherjee:2005wg}; as this algorithm is stochastic, in order to take into account possible statistical noise, we run it $\sim 100$ times obtaining a distribution of values from which we extract the best value of the evidence as the median of the distribution. The Evidence, $\mathcal{E}$, is defined as the probability of the data $D$ given the model $M$ with a set of parameters $\boldsymbol{\theta}$, $\mathcal{E}(M) = \int\ \mathrm{d}\boldsymbol{\theta}\ \mathcal{L}(D|\boldsymbol{\theta},M)\ \pi(\boldsymbol{\theta}|M)$, where $\pi(\boldsymbol{\theta}|M)$ is the prior on the set of parameters, normalized to unity, and $\mathcal{L}(D|\boldsymbol{\theta},M)$ is the likelihood function.

Once the Bayesian Evidence is calculated, one can obtain the Bayes Factor, defined as the ratio of evidences of two models, $M_{i}$ and $M_{j}$, $\mathcal{B}^{i}_{j} = \mathcal{E}_{i}/\mathcal{E}_{j}$. If $\mathcal{B}^{i}_{j} > 1$,  model $M_i$ is preferred over $M_j$, given the data. We have used the $\Lambda$CDM model, separately for both values of the pivot redshift we have defined above, as reference model $M_j$.

Even if the Bayes Factor $\mathcal{B}^{i}_{j} > 1$, one is not able yet to state how much better is model $M_i$ with respect to model $M_j$. For this, we choose the widely-used Jeffreys' Scale \cite{Jeffreys98}. In general, Jeffreys' Scale states that: if $\ln \mathcal{B}^{i}_{j} < 1$, the evidence in favor of model $M_i$ is not significant; if $1 < \ln \mathcal{B}^{i}_{j} < 2.5$, the evidence is substantial; if $2.5 < \ln \mathcal{B}^{i}_{j} < 5$, is strong; if $\ln \mathcal{B}^{i}_{j} > 5$, is decisive. Negative values of $\ln \mathcal{B}^{i}_{j}$ can be easily interpreted as evidence against model $M_i$ (or in favor of model $M_j$). In \cite{Nesseris:2012cq}, it is shown that the Jeffreys' scale is not a fully-reliable tool for model comparison, but at the same time the statistical validity of the Bayes factor as an efficient model-comparison tool is not questioned: a Bayes factor $\mathcal{B}^{i}_{j}>1$ unequivocally states that the model $i$ is more likely than model $j$. We present results in both contexts for readers' interpretation.

\section{Results of the observational tests}\label{sec:results}

The complete set of free parameters in our analysis is ${\Omega_m, \Omega_b, h, f_{i}, \alpha, \beta, \mathcal{M}_{B}, \Delta_{m}}$, where $f_{i}$ are the parameters corresponding to the various polynomial orders we have considered for each of the proposed parametrization of $f(z)$. We will focus in our comments only on the matter density parameter $\Omega_{m}$ and on the $f_{i}$ parameters, given that the other parameters are fully limited by imposed priors (e.g., $\Omega_{b}$ and $h$), or are independent of the cosmological background (e.g., SNeIa parameters, $\alpha, \beta, \mathcal{M}_{B}, \Delta_{m}$). In Tables  \ref{resultados1} and \ref{resultados2} we report the results obtained for such parameters in terms of the corresponding confidence levels, for the two different choices of the pivot redshift we have described in the previous section, i.e., $z_{piv}=10$ and $z_{piv}=100$. We also show the values of the Bayesian Evidence ratios, as defined in the previous section.

\subsection{Model 1: $\mathbf{\Lambda CDM}$}
First, let us concentrate on the results for $\Lambda$CDM. As we have shown in previous sections, for $\Lambda$CDM, not only $f(z)$ depends only on $f_{0}$ and $f_{3}$, but we also have the further conditions $f_{0}= 6(1-\Omega_{m}-\Omega_r)$ and $f_{3} = 3\Omega_{m}$. Note that we have always left $f_{0}$ and $f_{3}$ free in our MCMC analysis, so that it is interesting to check if the previous conditions are ``automatically'' verified by the $\Lambda$CDM case.

A straightforward check shows that the model (1), corresponding to $\Lambda$CDM, really satisfies Eq.~(\ref{fl}): from $\Omega_m$, we can calculate $3\Omega_{m} = 0.93^{+0.05}_{-0.03}$ for $z_{piv}=10$ and $3\Omega_{m} = 0.89^{+0.02}_{-0.02}$ for $z_{piv}=100$, which perfectly agrees with the corresponding free estimations of $f_{3}$. We can also calculate $6(1-\Omega_m-\Omega_r)=4.14^{+0.06}_{-0.09}$ for $z_{piv}=10$ and $6(1-\Omega_m-\Omega_r)= 4.21^{+0.04}_{-0.04}$ for $z_{piv}=100$, which also agree with our free estimations of $f_{0}$, mainly because of the larger errors on this parameter than on $\Omega_{m}$.

Before discussing generalizations and/or deviations from $\Lambda$CDM, let us note that the value of $\Omega_m$ does not really change from one case to another; there is tiny trend toward smaller values for $z_{piv}=100$ than for $z_{piv}=10$, but all the values are perfectly consistent at $1\sigma$ level.

\subsection{Models 2-4}
Now, considering case by case, let us consider model (2), which generalizes $\Lambda$CDM by adding intermediate powers corresponding to the $f_{1}$ and $f_{2}$ parameters. Given the very good agreement of $\Lambda$CDM with cosmological background data, we expect small deviations from it, if any. Actually, we can see how $f_{1}$ and $f_{2}$ are $\mathcal{O}(10^{-2})-(10^{-3})$ at $1\sigma$ confidence level. What is even more important, is that they are also consistent with zero, at $1\sigma$ for $z_{piv}=10$ and maximum at $2\sigma$ for $z_{piv}=100$. Thus, we can conclude that any $f(z)$ with this form is basically indistinguishable from $\Lambda$CDM at the present stage of observational data.

We also note that the simultaneous presence of both $f_1$ and $f_{2}$ seems to be tied to some degeneracy between the two parameters. In fact, for the models (3) and (4), where only one of the two parameters is present, the corresponding likelihoods are much more regular and far less noisy than in the case of model (2).

Anyway, we also stress that models (2), (3) and (4) have in common the presence of a constant term $f_0$ which, in some way, can bias the possible detection of any departure from $\Lambda$CDM because it can always be considered as representing a cosmological constant (on the limit $z\rightarrow0$). Actually, note that also for models (2)~(3) and (4) the same relations we have described for the $\Lambda$CDM case still hold. Eventually, a departure or agreement for them could tell us about the effective reliability and weight of the terms $f_{1}$ and $f_{2}$. In particular, we find that the relation $3\Omega_{m} = f_{3}$ is always perfectly satisfied by all models for both the pivot redshift values we have considered. The same holds true for the condition $6(1-\Omega_m-\Omega_r)=f_{0}$, except for model (2), which exhibits a value of $f_{0}$ clearly different from all the other cases, even if with larger errors.

Thus, we are not in the position to establish if the role of $f_{1}$ and $f_{2}$ is really effective and/or needed, or not: such relations should not hold in models (2)~(3) and (4), because for them $f(z) \neq f_{\Lambda}(z)$, but we find a good agreement, so that one might be led to think we are not actually detecting any deviation from the $\Lambda$CDM model. Moreover, the new parameters $f_{1}$ and $f_{2}$ are very small so that their role is really less significant than those of $f_{0}$ and $f_{3}$. But we cannot avoid to comment that these results strongly depend on the present observational status; future more precise data could help to discriminate between one model or another.

\subsection{Models 5-8}
More interesting are, from some point of view, models from (5) to (8) which explicitly avoid the introduction of the constant term $f_{0}$ so that they are not reducible to $\Lambda$CDM in any way. On the one hand, the first point to note is that, even in these cases, the parameters $f_{1}$ and $f_{2}$, when included, must be very small in the light of observations, ranging from $\mathcal{O}(10^{-2})$ to $\mathcal{O}(10^{-7})$. On the other, we see that the parameters corresponding to the lowest order powers $1/2$ and $1/4$ are very well constrained and the likelihood profiles have a very regular Gaussian shape. Furthermore, they seem to provide an equally good fit for the data with respect to $\Lambda$CDM with the same number of theoretical parameters.

Thus, the point to be understood here is: are these low order terms a real possible alternative to the cosmological constant, or should we consider them as a ``smeared'' version of the standard case only due to the low (for such type of discrimination) accuracy of the data? In fact, we have to consider that the power of the redshift we are considering might be low enough so that, due to the observational constraints we are using, they are actually mimicking the effective behaviour of a constant.

A possibly obvious trend is that the higher the order, the smaller the value of the corresponding parameter $f_{i}$ is (seen by comparing $f_0$ to $f_{1/4}$ and $f_{1/2}$); but this trend might be expected, given that $f_{1/2}$ and $f_{1/4}$ are by construction coupled to redshift dependent terms which could compensate the magnitude value with time dependence.

\subsection{Bayesian Evidence}
As we have mentioned before, apart from the information one can extract from the values of the parameters which best agree with the data, we have also computed the Bayes factor for each one of our proposed models against $\Lambda$CDM, the reference model. This statistical tool should provide us information about the reliability of each model. In general, we can see how $|\ln \mathcal{B}_{\Lambda}^{i}|<1$ so that there is no statistical significant preference for one model with respect to another.

\section{Summary and Conclusions}\label{sec:conclusions}

In this work we have introduced a different approach to convert general $f(R)$ theories in $f(z)$ models, which should be more easily connected to observational data and thus could shed some light on the explanation of the accelerated expansion of the universe and provide a more straightforward formulation to perform tests against observations.

The most attractive scenario would be to be able to solve the Friedmann equations analytically to find $H(z)$ solutions by proposing some $f(z)$ or viceversa. However, as we have mentioned in the introduction, this formulation is not possible in general, neither when studying the problem in terms of the Ricci scalar nor in terms of the redshift.

Then, what we have done has been to start from some $f(z)$ ans\"{a}tze and perform the analysis numerically. By studying some of the most known dark energy models we have been able to find some general potentially interesting features 
in order to shed some light on the expression of our proposals and restrict their arbitrariness by imposing some physically well-sounded and expected trends. In particular, we have found that a simple algebraic expression of a polynomial including just a constant and a third-order term can describe the general behaviour whatever $f(R)=f(z)$ model is expected to have in order to mimic $\Lambda$CDM at both high and low redshift. We need to stress that this polynomial does not need to include powers higher than order  three, as we have seen by analysing different dark energy models. Even if such analysis cannot be exhaustive, we know that $\Lambda$CDM works fine with most of the data we have and we expect that, if any, deviations from it might be small.

Thus, we have selected eight proposals varying the number of free parameters in order to analyze their viability according to the data. In particular, we have included middle-order terms in between the constant and the third-order one, checking for their compatibility with observations and their statistical soundness. And we have also proposed models with no constant term at all, so that they cannot be reduced at a $\Lambda$CDM scenario in any way.

A general conclusion we can extract from this work is that there are some f(z) polynomial models which are competitive with $\Lambda$CDM at the background level. We were especially interested in models (5) and (7) because they explicitly do not include a $\Lambda$-like term and, in fact, we have evidence indicating that these models are as reliable as $\Lambda$CDM when it comes to analyse observational data. Even though the Bayesian Evidence does not claim any significant difference with respect to $\Lambda$CDM, we still think this is an interesting result, which could be a useful guide when formulating and studying manageable alternative models of gravity. Moreover, we want to stress and keep in mind that $f(z)$ theories will probably not be the definitive answer to explain why the universe is expanding at an accelerated rate; but that was not really our objective here. Our goal was rather to provide a different perspective on how to relate the theoretical proposal of a modified gravity with observationally related quantities, and how to extract any kind of useful information which might contribute to the development of observationally reliable theories of gravity.

{\renewcommand{\tabcolsep}{1.5mm}
{\renewcommand{\arraystretch}{1.5}
\begin{table*}[ht!]
\caption{Results for $z_{piv}=10$.}\label{resultados1}
\begin{minipage}{0.95\textwidth}
\centering
\resizebox*{\textwidth}{!}{
\begin{tabular}{c|ccccccc|cc}
model & $\Omega_m$ & $f_0$ & $f_{1/4}$ & ${f_{1/2}}$ & ${f_1}~(10^{-2})$ & $f_2~(10^{-2})$ & $f_3$ & $\mathcal{B}^{i}_{\Lambda}$ & $\ln \mathcal{B}^{i}_{\Lambda}$ \\
\hline
1&$0.310_{-0.011}^{+0.015}$& $4.60_{-0.85}^{+0.63}$&-&-&-&-&$0.93_{-0.03}^{+0.05}$&1&0\\
2&$0.314_{-0.012}^{+0.012}$&$5.06_{-1.74}^{+1.75}$&-&-&$0.5^{+1.6}_{-1.0}$&$-1.1_{-1.8}^{+2.3}$&$0.94^{+0.03}_{-0.04}$&0.95&-0.05\\
3&$0.314_{-0.011}^{+0.015}$&$4.04_{-0.99}^{+0.87}$& -& -&-&$0.6^{+0.4}_{-1.1}$&$0.94^{+0.04}_{-0.03}$&1.05&0.05\\
4&$0.313_{-0.012}^{+0.015}$&$4.43_{-0.95}^{+0.62}$&-&-&$0.02_{-0.05}^{+0.14}$&-&$0.94^{+0.04}_{-0.04}$&1.23&0.21\\
5&$0.312_{-0.009}^{+0.013}$&-&-&$1.00^{+0.15}_{-0.19}$&-&-&$0.94^{+0.04}_{-0.03}$&1.11&0.10\\
6&$0.313_{-0.013}^{+0.012}$& -&-&$0.99^{+0.17}_{-0.17}$&$0.05_{-0.05}^{+0.04}$&$0.001^{+0.002}_{-0.001}$&$0.94_{-0.04}^{+0.04}$&0.91&-0.10\\
7&$0.314_{-0.013}^{+0.015}$&-& $1.91^{+0.34}_{-0.39}$&-&-&-&$0.94^{+0.05}_{-0.04}$&1.07&0.06\\
8&$0.313_{-0.012}^{+0.015}$&-& $1.91^{+0.31}_{-0.38}$&-&$0.09_{-0.07}^{+0.09}$&$\left(-0.09_{-0.19}^{+0.19}\right)\cdot10^{-3}$&$0.94^{+0.04}_{-0.04}$&1.06&0.06\\
\end{tabular}}
\label{resultados}
\end{minipage}
\end{table*}}}

{\renewcommand{\tabcolsep}{1.5mm}
{\renewcommand{\arraystretch}{1.5}
\begin{table*}[ht!]
\caption{Results for $z_{piv}=100$.}\label{resultados2}
\begin{minipage}{0.95\textwidth}
\centering
\resizebox*{\textwidth}{!}{
\begin{tabular}{c|ccccccc|cc}
model &$\Omega_m$&$f_0$& $f_{1/4}$ & ${f_{1/2}}$& ${f_1}~(10^{-2})$&$f_2~(10^{-2})$&$f_3$& $\mathcal{B}^{i}_{\Lambda}$ & $\ln \mathcal{B}^{i}_{\Lambda}$ \\
\hline
1&$0.298^{+0.006}_{-0.006}$& $4.44_{-2.40}^{+2.37}$&-&-&-&-&$0.89^{+0.02}_{-0.02}$&1&0\\
2&$0.313_{-0.015}^{+0.012}$& $6.00^{+2.72}_{-3.14}$&-&-&$-4.5^{+16.8}_{-30.5}$&$-2.6^{+2.0}_{-2.9}$&$0.94^{+0.04}_{-0.04}$&$0.86$&$-0.15$\\
3&$0.303_{-0.008}^{+0.010}$&$4.46_{-2.95}^{+2.00}$& -& -&-&$0.07^{+0.76}_{-0.84}$&$0.91^{+0.03}_{-0.02}$&1.09&0.09\\
4&$0.298_{-0.007}^{+0.007}$&$4.06_{-3.01}^{+2.64}$&-&-&$-0.002_{-0.001}^{+0.002}$&-&$0.89^{+0.02}_{-0.02}$&0.90&-0.10\\
5&$0.301_{-0.006}^{+0.006}$&-&-&$1.51^{+0.75}_{-0.69}$&-&-&$0.90^{+0.02}_{-0.02}$&1.24&0.22\\
6&$0.300_{-0.007}^{+0.007}$&-&-&$1.45^{+0.81}_{-0.94}$&$-3.5_{-1.4}^{+2.9}$&$-0.03^{+0.03}_{-0.01}$&$0.90^{+0.02}_{-0.02}$&1.01&0.008\\
7&$0.299_{-0.006}^{+0.006}$&-& $2.70^{+0.97}_{-1.07}$&-&-&-&$0.90^{+0.02}_{-0.02}$&1.12&0.11\\
8&$0.298_{-0.006}^{+0.006}$&-& $2.07^{+0.66}_{-0.65}$&-&$-0.007^{+0.004}_{-0.008}$&$\left(-0.013^{+0.008}_{-0.011}\right)\cdot10^{-3}$&$0.90^{+0.02}_{-0.02}$&1.08&0.07\\
\end{tabular}}
\end{minipage}
\end{table*}}}

\section*{Acknowledgments}
R.L. and M.O.B. were supported by the
Spanish Ministry of Economy and Competitiveness through
research projects No. FIS2014-57956-P (comprising FEDER
funds) and also by the Basque Government through research
project No. GIC17/116-IT956-16. M.O.B. acknowledges financial
support from the FPI grant BES-2015-071489. V.S. is funded by the Polish National Science Center Grant No. DEC-2012/06/A/ST2/00395.
This article is based upon work from COST Action CA15117
(CANTATA), supported by COST (European Cooperation
in Science and Technology).

\bibliography{biblio}{}

\end{document}